\documentstyle[aps,prb,twocolumn,floats]{revtex}

\begin{document}
\bibliographystyle{prsty}   
\input epsf
\draft
\wideabs{
\title {Heat transport by lattice and spin excitations in the
spin chain compounds SrCuO$_{2}$ and Sr$_{2}$CuO$_{3}$}

\author {A. V. Sologubenko, K. Giann\`{o}, H. R. Ott}
\address{Laboratorium f\"ur Festk\"orperphysik, ETH H\"onggerberg,
CH-8093 Z\"urich, Switzerland}

\author{A. Vietkine, A. Revcolevschi}
\address{Laboratoire de Physico-Chimie des Solides, Universit\'e Paris-Sud,
91405 Orsay, France}

\date{\today}
\maketitle

\begin{abstract}
We present the results of measurements of the thermal conductivity 
 of the quasi
one-dimensional spin S=1/2 chain compound  SrCuO$_{2}$ in
the temperature range between 0.4 and 300~K 
along the directions parallel and perpendicular to 
the chains. An
anomalously enhanced thermal conductivity is observed along the chains. 
The analysis of the 
present data and a comparison with analogous recent results for 
Sr$_2$CuO$_3$ and other similar 
materials demonstrates
that this behavior is generic for  cuprates with copper-oxygen chains 
and strong intrachain interactions.
The observed anomalies are attributed to the
one-dimensional energy transport by spin excitations (spinons), 
limited by the
interaction between spin and lattice excitations. The energy 
transport along the spin chains has a non-diffusive character, in 
agreement with theoretical predictions for integrable models. 
\end{abstract}
}    

\section{Introduction}

The physics of one-dimensional (1D) magnetic systems has been of interest for 
some time but particularly significant progress, both in experiment 
and theory, has been  
made during the last several years. The related research activities 
have been growing because the quantum nature of  
low-dimensional low-spin systems promises
a rich variety of phenomena to be explored. Among other features, 
the transport of energy in 1D magnetic systems is expected to 
be highly unusual. 
A number of models describing one-dimensional
systems are integrable, implying, for instance, conservation of 
energy current and, as a consequence, ballistic energy propagation and 
divergent thermal conductivity. 
The question whether the energy transport is diffusive or ballistic 
is currently under active discussion for atomic\cite{Giardina2000,Gendelman2000,Hu2000} 
as well as spin\cite{Castella95,Saito96,Zotos97} 1D systems.

The 1D Heisenberg  $S$=1/2  systems with antiferromagnetic coupling 
between adjacent spins are of particular interest. It is well 
established that elementary excitations in such systems are not 
$S$=1 magnons but $S$=1/2 quantum solitons,\cite{Faddeev81} commonly called spinons.
Spinons can only be created or destroyed in pairs but they may be 
treated as 
free particles, at least at large spatial separation.\cite{McRae98,Bernevig2000}
The interaction of spinons, as quantum solitons, with structural defects and other 
quasiparticles, such as 
phonons, is poorly investigated both theoretically and experimentally, 
although in classical Heisenberg and Ising 1D magnetic 
systems, the  interaction of solitons with defects, phonons, and magnons 
demonstrates a number of interesting 
features\cite{Schoubinger81,Buijs82,Wysin87,deGronckel91,Evangelou96} 
(for a review, see Ref.~\onlinecite{Baryakhtar92}). 

In order to probe the features of energy transport 
in $S$=1/2 Heisenberg chains, only a limited number of 
experiments, such as inelastic neutron 
scattering\cite{Racz2000} or NMR,\cite{Zotos99} is available.   
In case of a sufficiently strong coupling between the phonon and 
the spin subsystems, the energy transport mediated by magnetic excitations 
can also be monitored via experiments probing the thermal conductivity. 
Although the spin-lattice interaction as well as the influence of defects and 
interchain interactions make the spin system non-integrable,  
an anomalous behavior of the spin-mediated thermal conductivity may, 
nevertheless, be expected.
Establishing the heat transport carried by spin excitations may, in 
addition, reveal information concerning 
magnetic defects and the spin-lattice interaction. The 
spin-lattice interaction 
is very  important in 1D magnetic system, because it leads to  
modifications of the spectrum of spin excitations\cite{Sushkov99} and,
under special circumstances, to the formation of new phases, such as  
the spin-Peierls dimerised state. Besides defects and phonons, interchain interactions 
are also anticipated to influence the energy transport in the spin 
system and hence may be probed by thermal-conductivity 
experiments.   

In previous work  concerning the thermal conductivity $\kappa(T)$ in quasi-1D 
$S$=1/2 Heisenberg systems,  observations of 
some spin-mediated heat transport, in addition to the dominant phonon 
contribution, have been reported for  KCuF$_{3}$,\cite{Miike75} 
CuGeO$_{3}$,\cite{Takeya2000} 
and Yb$_{4}$As$_{3}$.\cite{Koeppen99}
The most pronounced 
effects, however, have been observed  in 
(Sr,Ca)$_{14}$Cu$_{24}$O$_{41}$, containing double-leg Cu-O ladders  
with  180$^\circ$ Cu-O-Cu bond angles.\cite{Sologubenko2000_ladder,Kudo01JPSJ} 
The thermal conductivity along the ladder direction exhibits 
an anomalous double-peak temperature dependence.
It has fairly well been established by a detailed 
analysis\cite{Sologubenko2000_ladder}
that this two-peak feature in $\kappa(T)$ of
(La,Sr,Ca)$_{14}$Cu$_{24}$O$_{41}$  is caused by two main 
contributions, one due to phonons,
responsible for the low-temperature peak,  and  the other due to 
itinerant spin excitations in
the ladders, causing the additional high-temperature maximum.

\begin{figure}[t]
 \begin{center}
  \leavevmode
  \epsfxsize=0.9\columnwidth \epsfbox {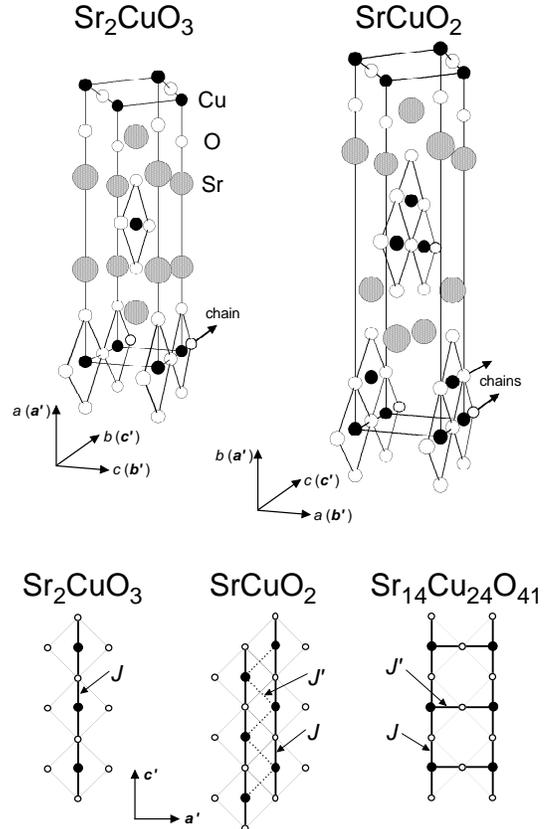}
   \caption{
  The schematic unit cells of the crystal structures of SrCuO$_{2}$ and  
  Sr$_{2}$CuO$_{3}$. For each material, the layout of the Cu-O chains 
  is emphasized. For comparison, the structure of the ladder in 
  Sr$_{14}$Cu$_{24}$O$_{41}$ is also shown.
  }
\label{Structures} 
\end{center}
\end{figure}

In order to test the above mentioned conjectures concerning 
energy transport by spin 
excitations in $S$=1/2 1D Heisenberg systems, we have investigated the 
low-temperature thermal conductivities of  two 
materials containing 180$^\circ$ Cu-O chains, namely SrCuO$_{2}$  and
Sr$_{2}$CuO$_{3}$.  
The crystallographic unit cells\cite{Teske69,Teske70} of the two compounds are 
schematically drawn in Fig.~\ref{Structures}.
The intrachain exchange coupling $J$ of 2100--3000~K\cite{Ami95,Suzuura96,Motoyama96,Johnston97}
in these two materials is  of the same order of magnitude as the interactions 
along the ladder legs 
in  Sr$_{14}$Cu$_{24}$O$_{41}$  ($J$=1510~K).\cite{Eccleston98}
The compounds considered here exhibit different arrangements of the 
same type of chains. 
The structure of Sr$_{2}$CuO$_{3}$ contains Cu-O chains with a very small interchain 
interaction $J'$ ($\alpha \equiv J'/J \sim 10^{-5}$).\cite{Motoyama96}
SrCuO$_{2}$ is built 
by double Cu-O chains forming Cu-O ribbons containing Cu-Cu zig-zag chains (see 
Fig.~\ref{Structures}). The interaction between the two chains occurs 
via 90$^\circ$ Cu-O-Cu bonds providing a weak 
($\left| \alpha \right| = $ 0.1-0.2) ferromagnetic 
interaction.\cite{Rice93} 
For comparison, the chains forming the double-leg ladders 
in (Sr,Ca)$_{14}$Cu$_{24}$O$_{41}$  are connected via  180$^\circ$ 
Cu-O-Cu bonds (``rungs''), therefore for the ladders $\alpha = 0.55$ is 
rather large.\cite{Eccleston98} The spin excitation spectrum for the 
single-chain compound Sr$_{2}$CuO$_{3}$ is 
gapless. Because the  
interaction between the chains forming the Cu-O ribbons in the SrCuO$_{2}$ is ferromagnetic, 
the excitation spectrum remains gapless,\cite{White96} in contrast to the 
gapped spectrum of spin ladders where the interaction between the chains is 
antiferromagnetic. 

In this paper, we present new experimental results on the thermal 
conductivity  of SrCuO$_{2}$  along 
the main crystallographic directions in the temperature region between 
0.4 and 300~K. 
These data are analyzed together with those 
for Sr$_{2}$CuO$_{3}$ measured and presented previously.\cite{Sologubenko2000_213} 
We offer a detailed analysis of phonon and spinon contributions to 
the thermal conductivity along different crystallographic directions. 
We show that the thermal transport 
perpendicular to the chain direction is predominantly phononic at all 
temperatures, whereas along the 
chain direction the spinon contribution is dominant at elevated 
temperatures. 
The scattering of phonons is mainly via other
phonons, at dislocations, and at sample boundaries. The spinon thermal 
conductivity may reliably be evaluated only in the high temperature 
region above 60~K where it is predominantly limited by the 
spin-lattice interaction. 
At very low temperatures, the spin-lattice interaction appears to be 
rather weak and therefore the intrinsic 
spinon thermal conductivity cannot be observed. 

For clarity we introduce an auxiliary notation of crystallographic 
directions. For both compounds we denote the 
direction along the chains as $c'$ ($c$ for 
SrCuO$_{2}$  and $b$ for Sr$_{2}$CuO$_{3}$), the 
direction perpendicular to the CuO squares as $b'$ ($a$ for 
SrCuO$_{2}$  and $c$ for Sr$_{2}$CuO$_{3}$), and the direction in the 
CuO square plane but perpendicular to the chains as $a'$ (see  
Fig.~\ref{Structures}). This notation is compatible  with the 
crystallographic axes of 
the spin-ladder system (Sr,Ca)$_{14}$Cu$_{24}$O$_{41}$, for which 
$a'$, $b'$, and $c'$  correspond to the $a$, $b$, and $c$ directions, respectively. 

\section{Experimental details}

The single crystals of SrCuO$_{2}$ and  Sr$_{2}$CuO$_{3}$ were grown 
by a traveling-solvent floating-zone method.\cite{Revcolevschi97} The high 
structural quality of 
the crystals was confirmed by neutron diffraction experiments.\cite{Revcolevschi97}

From each crystal, three specimens in the form of rectangular bars 
with typical dimensions 2.5$\times$1$\times$1 mm$^{3}$ and with the longest 
dimension parallel to each principal axis were cut with a thin-blade 
diamond saw.  The thermal 
conductivity was measured by a standard steady-state heat-flow 
technique, fixing the sample at one end to a heat sink. 
The temperature gradient along the sample was produced by 
a heater (1~k$\Omega$  chip resistor) glued to the opposite end of 
the sample, and monitored in 
overlapping temperature ranges by a matched pair of RuO$_{2}$ thermometers 
at $T < 5$~K  and by differential Chromel-Au + 0.07\% Fe thermocouples at higher 
temperatures. The temperature difference between the thermometers was 
typically 1.5\% of the absolute mean temperature of the sample. 
The estimated accuracy of $\pm$ 10\% of the absolute value of 
$\kappa(T)$ is mainly caused by the uncertainty 
of the sample geometry, the relative error being of the order of 0.5\% 
of the measured values of $\kappa$.  

\section{Experimental results}

\begin{figure}[t]
 \begin{center}
  \leavevmode
  \epsfxsize=0.9\columnwidth \epsfbox {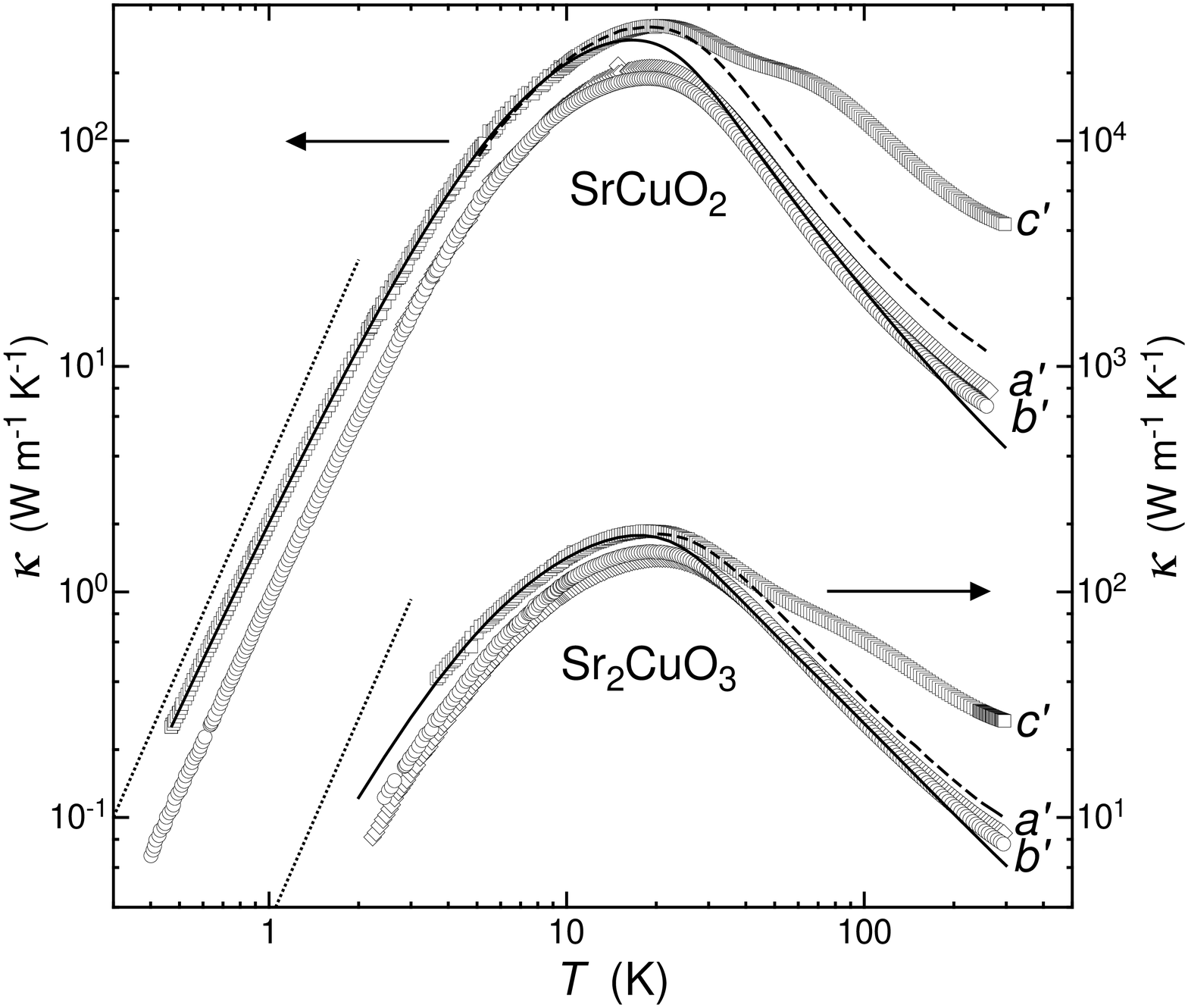}
   \caption{
  Temperature dependences of the thermal conductivities of SrCuO$_{2}$
  and Sr$_{2}$CuO$_{3}$  along the $a'$, $b'$, and
$c'$ axes. The dotted lines represent 
estimated limits of the thermal conductivity due 
to the finite size of the samples. The solid and dashed lines 
represent different evaluations of the phonon  contribution to 
$\kappa_{c'}$ as described in the text.
  }
\label{Kappa} 
\end{center}
\end{figure}

The results of the thermal conductivity measurements for SrCuO$_{2}$ are presented in
Fig.~\ref{Kappa}, together with our previous results\cite{Sologubenko2000_213} on the
thermal
conductivity of Sr$_{2}$CuO$_{3}$.
 For both materials $\kappa(T)$ reveals similar general features. 
In particular, the thermal conductivities along the directions perpendicular  
to the chains ($\kappa_{a'}$, $\kappa_{b'}$) are similar not
only in their features of the temperature dependence, but even in absolute
values. At the same time, the thermal
conductivity along the chains is distinctly higher than $\kappa(T)$
perpendicular to the chains in both cases. This is particularly 
evident at temperatures above the $\kappa(T)$ maxima.  
At room temperature, this excess conductivity for the double-chain 
material SrCuO$_{2}$ 
exceeds that for the single-chain compound Sr$_{2}$CuO$_{3}$ by almost a 
factor of two.
In the temperature region between  1 and  6~K, where  magnetic phase transitions have been 
reported for both materials,\cite{Keren93,Kojima97,Matsuda97,Zaliznyak99} 
no anomalous features in $\kappa(T)$ have been observed. 
Since these transitions are indicated by anomalies of the 
magnetic specific heat,\cite{Sologubenko2000_213,Matsuda97}   
the interaction between the spin and the lattice systems as well 
as the spin contribution to the measured thermal conductivity must be 
negligible in this temperature region for both materials.

Because both SrCuO$_{2}$ and Sr$_{2}$CuO$_{3}$ are electrical
insulators, electrons do not participate in the heat transport. 
Below, we thus analyze the experimental 
data assuming that the main heat carriers are phonons and spin 
excitations.

\section{Analysis of the phonon contribution}\label{sPhononContribution}

\subsection{The model and fitting procedure}\label{ssModel}

In general, the behavior of the phonon thermal conductivity
$\kappa_{\rm ph}(T)$ is determined by scattering of phonons by 
structural defects, phonons and other quasiparticles. 
At temperatures $T\ge \Theta_{D}$, where $\Theta_{D}$ is the
Debye temperature, the phonon-phonon Umklapp-processes (U-processes) usually 
dominate and it often turns out that 
$\kappa_{\rm ph}\propto T^{-\alpha}$, where $\alpha \sim
1$. The mean free path of the phonons is very small and comparable to
interatomic distances.
At temperatures below $\Theta_{D}$ the probability of  U-processes
decreases exponentially and the phonon mean free path increases 
accordingly.
The maximum mean free path of phonons, normally reached only at very 
low temperatures, is limited by the size of a sample. At this, so 
called Casimir limit, the thermal conductivity  may be calculated by 
$\kappa _{\rm ph}=1/3 C_{\rm ph} v L_{b}$, where $C_{\rm ph}$ 
is the lattice specific heat, $v$ is the average sound velocity along 
the chosen direction, and $L_{b}$ 
is a constant related to the sample size. Hence, 
for isotropic materials,  $\kappa_{\rm ph} \propto T^{3}$ at low temperatures.
These two sources for scattering of phonons are always present in 
any material, generating some typical features of 
$\kappa_{\rm ph}(T)$ including a 
distinct maximum at  $T_{\rm max} \leq \Theta_{D}/10$. 
Various defects, such as impurities, dislocations, stacking faults etc., 
also participate in phonon scattering, thereby reducing the thermal conductivity, 
especially in the region of $T_{\rm max}$. Various quasiparticle 
excitations such as magnons, electrons, excitons etc. also
influence the thermal conductivity by scattering phonons and 
by providing additional channels of heat transport.    

In order to identify the 
dominant mechanisms of phonon scattering and to estimate their relative 
importance, we have fitted the experimental data by using the Debye 
approximation 
of the phonon spectrum combined with the relaxation-time approximation
for calculating the thermal conductivity. The model assumes the same group 
velocities and relaxation 
rates for phonons of different polarizations. A more accurate 
and more complicated analysis, taking into 
account possible differences in the scattering rates for phonons of different 
polarisations, would require a detailed knowledge of elastic constants, which is still missing.  
The phonon thermal conductivity is thus 
calculated as
\begin{equation}\label{eLambda}
    \kappa_{\rm ph} =  \frac{ k_{\rm B}}{ 2 {\pi}^{2} v } \left( 
    \frac{k_{\rm     B}}{\hbar} \right) ^{3} T^{3}  
    \int\limits_{0}^{\Theta_{D}/T}
    \frac{x^{4}e^{x}}{(e^{x}-1)^{2}}\tau (\omega,T) dx,
\end{equation}
where $\omega$ is the frequency of a phonon, $\tau(\omega,T)$ is the 
corresponding relaxation 
time, $\Theta_{D}$ is the Debye temperature, and $x=\hbar\omega/k_{\rm B} T$.  
As a further simplification we assume 
that all mechanisms of phonon scattering act independently.  In this 
case, 
\begin{equation}\label{eTau}
 \tau^{-1} = \sum \tau_{i}^{-1},  
\end{equation}
where each  term  $\tau_{i}^{-1}$ corresponds to an individual 
independent scattering mechanism. For instance, 
\begin{equation}\label{eTauB}
 \tau^{-1}_{b}= v / L_{b} 
\end{equation}
for phonon scattering by sample boundaries,
\begin{equation}\label{eTauPD}
 \tau^{-1}_{pd}=A \omega^{4} 
\end{equation}
for phonon scattering by point defects (Rayleigh scattering), and
\begin{equation}\label{eTauU}
 \tau^{-1}_{U} = B \omega^{2} T \exp\left( -\Theta_{D}/bT \right )
\end{equation}
for phonon-phonon U-processes.
As indicated 
in Eq.~(\ref{eTau}), there may be other contributions to $\tau^{-1}$, such as 
phonon-phonon N-processes, phonon-spin scattering, etc. 
In the first step only the three main 
mechanisms (boundaries, point defects and U-processes) were considered for all samples.
In those cases where the fitting of the data with 
a reasonable accuracy failed, 
additional terms representing relevant scattering processes were 
introduced in Eq.~(\ref{eTau}).

For the fitting procedure we employed the Levenberg-Marquardt algorithm. 
The quantities $L_{b}$, $A$, $B$, and $b$, defined in 
Eqs.~(\ref{eTauB}) to (\ref{eTauU}) 
 were treated as freely adjustable parameters.
The criterion of selecting additional terms in Eq.~(\ref{eTau}) was the 
minimization of the mean-square deviation $\chi^{2}$. 
We used the $\Theta_{D}$ values  
of 441~K for  Sr$_{2}$CuO$_{3}$ 
(Ref.~\onlinecite{Sologubenko2000_213}) and 357~K for  SrCuO$_{2}$ 
(Ref.~\onlinecite{MatsudaPrivate}).  The mean sound velocity  $v$ was  
calculated from the values of $\Theta_{D}$, using the equation 
$ v=\Theta _D\left( {{{k_B} \mathord{\left/ {\vphantom {{k_B} \hbar }} \right. 
\kern-\nulldelimiterspace} \hbar }} \right)(6\pi ^2n)^{-1/3}$, 
where $n$ is the number density of atoms.

\subsection{Phonon thermal conductivity perpendicular to the chain 
direction}\label{KappaPhab}

As may be seen in Fig.~\ref{Kappa}, 
the thermal conductivities of SrCuO$_{2}$ and Sr$_{2}$CuO$_{3}$ in the directions
perpendicular to the chains exhibit the 
temperature dependence that is typical for phonon heat transport. 
This includes  $\kappa\propto T^{-1}$ at high temperatures and 
approaching $\kappa\propto T^{3}$ at low temperatures, the latter 
being 
expected if the phonon mean free path $l_{\rm ph}$ is constant. 
Since $l_{\rm ph}$ turns out to be shorter than the smallest sample 
dimensions even at 0.4K (for SrCuO$_{2}$), 
the influence of  defects cannot be neglected.  

It turns out that the three mechanisms initially 
included in Eq.~(\ref{eTau}) were not enough to fit the data 
in the entire covered temperature range. 
However, after considering various other possible scattering processes,  
very good agreement  between the experimental and the
calculated $\kappa_{a'}(T)$ and $\kappa_{b'}(T)$ curves 
in the temperature region between 0.4 and 100~K was achieved  by
including two additional terms in Eq.~(\ref{eTau}).
The first one, 
\begin{equation}\label{eOmegaD}
   \tau^{-1}_{d} = C \omega,
\end{equation}
where $C$ is a fit parameter, is usually attributed to phonon scattering by strain fields of 
dislocations.\cite{Klemens58}  This term proved to be essential for 
fitting the temperature dependence of the
thermal conductivity at $T<T_{\rm max}$. 
The second term is 
of  resonant type\cite{Sheard73} 
\begin{equation}\label{eTauRes}
   \tau_{res}^{-1}  =
D{{\omega ^4} \over {(\omega ^2-\omega_0^2)^2}}\left( {1-c g^2(\omega 
_0,T)} \right),
\end{equation} 
where $\omega_0$ is the resonance frequency,  $c$ is the fractional 
concentration of scatterers and $D$ is a factor characterizing the 
strength of the resonant scattering.  The function $g(\omega 
_0,T) \equiv (p_{-} -p_{+})$, describing the difference in population of 
the upper ($p_{+}$) and lower ($p_{-}$) states was taken as
$g(\omega _0,T) = \tanh (\hbar\omega _0/2 k_B 
T )$,  corresponding to a two-level system with singlet upper and lower 
levels.  
The parameters $D$, $c$, and $\omega_0$ are again fit parameters.  A resonant term of 
this type was initially proposed for phonon scattering by magnetic 
impurities,\cite{Sheard73} but was also successfully used to explain the scattering 
of acoustic phonons by flat optical phonon modes.\cite{Wybourne85}

\begin{figure}[t]
 \begin{center}
  \leavevmode
  \epsfxsize=0.9\columnwidth \epsfbox {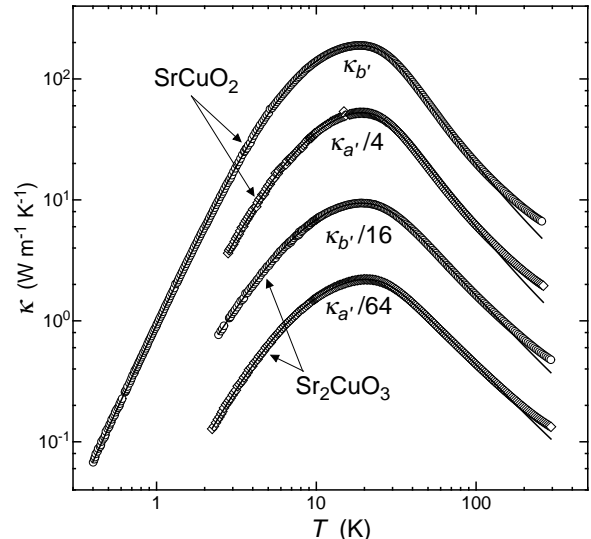}
   \caption{
  A comparison of the experimental (open circles and diamonds) 
  and calculated (solid 
  lines) $\kappa(T)$ along  the directions perpendicular to the chains. 
  For clarity, the curves for different samples are scaled in order 
  to achieve a reasonable separation between them.
  }
\label{LamFit} 
\end{center}
\end{figure}

The fit values of the 
adjustable parameters $L_{b}$, $A$, $B$, $C$, $D$, $\omega _0$, $c$, 
and $b$  are listed in Table~\ref{t_1}. The $\kappa(T)$ 
curves calculated using these values are shown in Fig.~\ref{LamFit}, 
together with the experimental data. 

For a sample with a  rectangular cross-section $S_{cs}$, a fully 
diffusive scattering of phonons at the sample boundaries leads to 
$L_{b}=2\sqrt {S_{cs}/\pi}$. The expected values of  $L_{b}$ compatible with 
the dimensions of our samples are of the order of 1~mm, i.e.,   
close to the fit values given in Table~\ref{t_1}.  The small values 
of $L_{b}$ may possibly be caused by large  
defects, such as microcracks present in the samples. 

The fitting procedure gave values of the parameter $A$ in Eq.~(\ref{eTauPD}) equal to 
0 or less than $10^{-45}$~s$^{3}$. With 
$A$ that small, the influence of 
point defects on the phonon thermal conductivity is negligible  
in comparison with other processes of phonon scattering.

For materials  with $\nu$ atoms in the unit cell, 
the parameter $b$ in Eq.~(\ref{eTauU}) 
is  expected to be given by $b\sim 2 \nu^{1/3}$ (Ref.~\onlinecite{Slack79}). 
The fit  values of $b$ presented in Table~\ref{t_1} are in reasonable agreement 
with this expectation.
The parameter $B$, characterizing the 
strength of U-processes, depends in a complex manner on elastic and 
anharmonic constants of a material. 

The parameter $C$, associated with 
dislocation scattering, is given by\cite{Klemens58}     
\begin{equation}\label{eDis}
C= k n_d\bar b^2\gamma ^2,
\end{equation} 
where $n_d$ is the number of dislocations per unit area, $\bar b$ is 
the Burgers vector, $\gamma$ is the Gr\"uneisen constant, and $k$ is 
a  constant of the order of $10^{-1}$. Assuming that $\gamma \sim 2$ 
and $b \sim 4$\AA, we get $n_d \sim 10^{14}$~m$^{-2}$.   

The origin of a resonant scattering of phonons 
in SrCuO$_{2}$ and Sr$_{2}$CuO$_{3}$ is not a priori clear.
However, the fact that $c$ defined in  Eq.~(\ref{eTauRes}) turns out 
to be close to 1 
means that the number of scattering centers is of the order of the number of 
unit cells in the sample, which severely limits the choice of scattering 
processes and excludes, for example, dilute magnetic 
impurities as scatterers. One possible origin of such scattering could 
be the interaction of acoustic phonons with a flat optical phonon 
branch, the situation that was promoted in Ref.~\onlinecite{Wybourne85}. 
Another possible source of resonant scattering are excitations of 
segments of spin chains of finite lengths, created by the inclusion of nonmagnetic defects.
In this latter case, the expression for  $g(\omega_0,T)$ in Eq.~(\ref{eTauRes}) 
should be changed to describe the degeneracy of the ground and excited 
states of a chain segment. For chain segments with an even 
number of spins, which are expected to prevail in $S=1/2$ AFM Heisenberg 
chains,\cite{Nishino2000} the low-energy resonance corresponds to 
a singlet-triplet transition and, therefore,  
$g(\omega_0,T) = (1-\exp(\hbar\omega_0/ k_B T )) /(1+3\exp(\hbar\omega_0/ k_B T ))$. 
It appears, however, that choosing between $g(\omega_0,T)$ valid for either 
a singlet-singlet or  a singlet-triplet excitation has only a minor effect 
on the resulting quality of the  fit and the values of the parameters in 
Eq.~(\ref{eTauRes}). Therefore, at this point, we cannot distinguish 
these two possibilities.       

As one can see in Fig.~\ref{LamFit}, the experimental data 
slightly deviate from the approximation above 100~K. The magnitude of the deviations 
increases with increasing temperature and, at 
room temperatures, it is as large as 1.5--2 W~(m~K)$^{-1}$. 
Experimentally, heat losses by radiation result in measured values of 
$\kappa (T)$ higher than the intrinsic thermal conductivity. 
This effect is often observed at high temperatures 
for samples with poor  thermal conductivity, if no special precautions are taken. 
Estimates have shown, however, that in our experimental setup the heat losses 
via radiation are at least an order of magnitude too small to explain 
the excess conductivity and hence 
the observed deviation should be attributed to an intrinsic mechanism.
Since with increasing temperature the heat is mainly transported  by 
high-frequency phonons for which the Debye model is not a good 
approximation, the deviations may in principle be attributed to shortcomings of 
the employed simple model. 
Another source of high-temperature deviations may be an
additional heat transport by optical phonons. 
We recall that high-temperature deviations of the same magnitude 
can also be discerned in our $\kappa_{a'}(T)$ data
for the spin-ladder system (Sr,Ca)$_{14}$Cu$_{24}$O$_{41}$.\cite{Sologubenko2000_ladder}
The similarity of the high-temperature behavior of $\kappa_{a',b'}(T)$ of spin-chain 
and spin-ladder materials seems intriguing and may hint to a common 
origin of the feature such as, as discussed later, diffusive energy 
transport via the spin system.  

\subsection{Phonon thermal conductivity parallel to the chain direction}

\begin{figure}[t]
 \begin{center}
  \leavevmode
  \epsfxsize=0.9\columnwidth \epsfbox {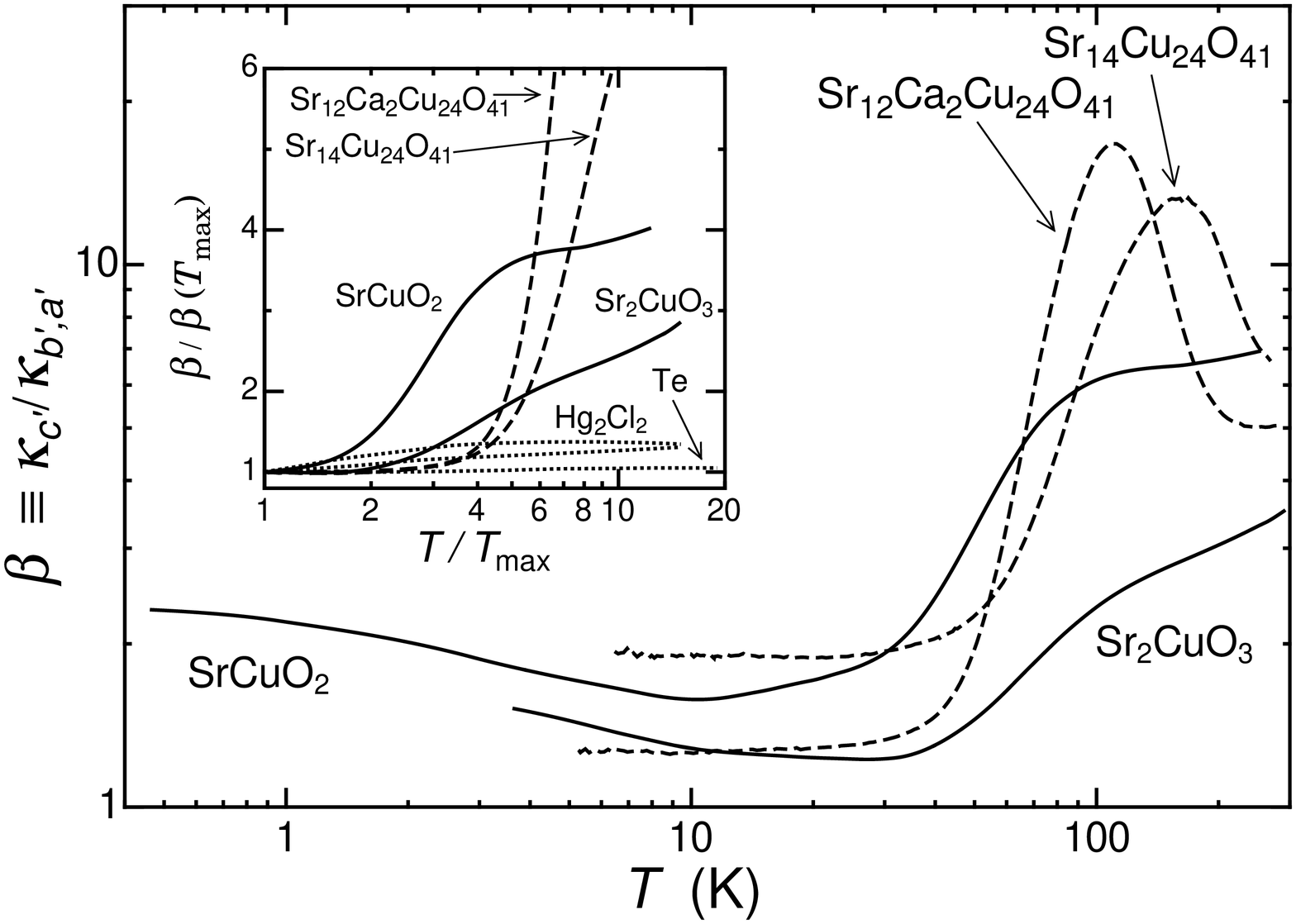}
   \caption{
  Temperature dependences of the anisotropy ratios 
  $\kappa_{c'}/\kappa_{b'}$ for SrCuO$_{2}$ and Sr$_{2}$CuO$_{3}$,  and  
  $\kappa_{c'}/\kappa_{a'}$ for (Sr,Ca)$_{14}$Cu$_{24}$O$_{41}$ (data 
  of Ref.~\protect\onlinecite{Sologubenko2000_ladder}).
   The inset illustrates the relative change of the anisotropies with 
   temperature above $T_{\rm max}$ for the same compounds and also for 
   Hg$_{2}$Cl$_{2}$ (data of Ref.~\protect\onlinecite{Koss90}) and 
   Te (data of Ref.~\protect\onlinecite{Adams67}). 
  }
\label{KAnisotropy} 
\end{center}
\end{figure}

In contrast to the successful interpretation of the thermal conductivity 
perpendicular to the chains  by considering phonon heat transport alone,  
we failed completely in obtaining a reasonably good fit of  
the anomalous  $T$-dependence of $\kappa_{c'}$ at 
temperatures above $T_{\rm max}$ with a similar approach.
In order to emphasize the anomalous feature in 
$\kappa_{c'}(T)$,  we show the 
anisotropy ratio $\beta=\kappa_{c'} /\kappa_{b'}$ in Fig.~\ref{KAnisotropy}. 
For comparison,  
$\beta=\kappa_{c'} /\kappa_{a'}$ ratios for the 
spin-ladder system Sr$_{14-x}$Ca$_{x}$Cu$_{24}$O$_{41}$ ($x$=0, 2) 
are included in Fig.~\ref{KAnisotropy}. 
At low temperatures the anisotropy is relatively 
weak and varies little with increasing temperature.
A distinct rise of $\beta$ is observed above 30-40~K.  

The obvious similarity of the three systems containing 
similar Cu-O chains  with different interchain interactions suggests a 
common cause for this anomaly. 
The anomaly might be due to drastic changes in the phonon heat 
transport above $T_{\rm max}$, but we argue below that this is not the case. 
Indeed, in the vast literature on properties of Sr$_{2}$CuO$_{3}$ 
and  SrCuO$_{2}$ no observation of any phase transition above 6~K has been 
reported and,
therefore, no drastic changes  in elastic interatomic interactions 
at $T > T_{\rm max}$ is expected. Without such a change affecting the lattice, the 
chain-like structure of the lattice itself cannot explain the strong 
temperature-dependent anisotropy of the thermal conductivity above $T_{\rm max}$.
To illustrate this point,  we show in the inset of 
Fig.~\ref{KAnisotropy} the anisotropy of the phonon thermal 
conductivities  
for Hg$_{2}$Cl$_{2}$  and Te (data 
from Refs.~\onlinecite{Koss90} and \onlinecite{Adams67}, 
respectively), 
two chain-type materials with pronounced elastic anisotropies.
Hg$_{2}$Cl$_{2}$ and Te were chosen as examples because 
in both cases $\kappa(T)$ varies as $T^{-1}$ at high 
temperatures, i.e., the phonon-phonon scattering dominates  above $T_{\rm max}$.
This demonstrates that even for chain-type materials with strong elastic 
anisotropies,  the anisotropy of the phonon thermal conductivity above T$_{\rm 
max}$ is only weakly temperature-dependent.

From the point of view  of 
elastic properties, the spin-chain cuprates are rather ordinary, weakly anisotropic 
materials. For example, the anisotropy of the sound velocities of 
Sr$_{14}$Cu$_{24}$O$_{41}$ is found to be very weak.\cite{Koenig97}
Therefore, as we have already argued in 
Ref.~\onlinecite{Sologubenko2000_213}, well above $T_{\rm max}$ one may 
expect not only a weak temperature dependence of $\beta$, but also 
a rather small anisotropy of the phonon-phonon scattering, and 
therefore $\beta\simeq 1$. 

\begin{figure}[t]
 \begin{center}
  \leavevmode
  \epsfxsize=0.9\columnwidth \epsfbox {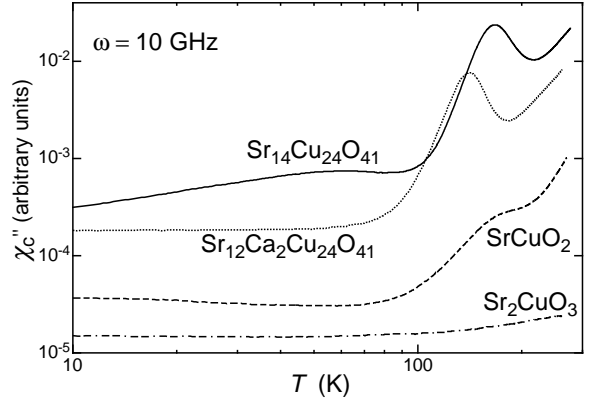}
   \caption{
  Imaginary part of the dynamic magnetic susceptibility 
  $\chi''_{c}({\rm 10GHz},T)$  (reproduced from 
  Ref.~\protect\onlinecite{Zhai99}), for the compounds that are 
  discussed here.}
\label{Lam-chi} 
\end{center}
\end{figure}

On the other hand, the high-temperature features of $\kappa_{c'}(T)$ 
do correlate with features in magnetic properties. 
In Fig.~\ref{Lam-chi} we reproduce 
the results of Zhai {\it et al.}\cite{Zhai99} for 
the dynamic magnetic susceptibilities  $\chi''_{c}(\omega,T)$ of 
SrCuO$_{2}$,  Sr$_{2}$CuO$_{3}$, and 
Sr$_{14-x}$Ca$_{x}$Cu$_{24}$O$_{41}$ ($x=0$, 2)
in the 
$c'$-direction, measured at $\omega =$ 10~GHz. 
By comparison of Figs.~\ref{KAnisotropy} and  \ref{Lam-chi},  
the correlation between the thermal 
conductivity and the magnetic susceptibility along the chain direction
is obvious\cite{_Comment1} and indicates that 
the anomalous $\kappa_{c'}(T)$ is of magnetic origin.

Based on the arguments presented above, 
the increase of $\beta(T)$ for SrCuO$_{2}$ and Sr$_{2}$CuO$_{3}$ with 
increasing temperature  is  
attributed to an additional, non-phononic channel of heat transport 
along the chain direction. Most naturally, this excess thermal 
conductivity is  associated with an energy transport 
via spin excitations of the spin chains. 
This additional heat transport is large enough to  be 
separated reliably from the phonon contribution.  

In order to achieve this separation of 
the spin-mediated contribution to the thermal conductivity from 
the experimental data on $\kappa_{c'}$, some assumptions about the phonon 
background have to be made. 
As we have argued before,\cite{Sologubenko2000_213}
the phonon-phonon scattering is most likely 
almost isotropic. For the evaluation of the relaxation rate via 
phonon-phonon scattering along the chain direction, 
one can therefore use the average values of the parameters  $B$ and $b$, obtained by 
fitting  $\kappa(T)$ along the other directions (see Table~\ref{t_1}).  
With this assumption, we have fitted the  data for 
$\kappa_{c'}(T)$ at low temperatures ($T\leq 10$~K) where 
the influence of spin-mediated heat transport, as argued above, is negligible. 
The fitting procedure was the same as for $\kappa_{a'}$ and 
$\kappa_{b'}$ but now, only the parameters $L$, $A$, $C$, and 
$D$ were kept free.  For the parameters $c$ (concentration of resonant 
scatterers) and $\omega_{0}$ (resonance frequency), again the average 
values obtained by fitting $\kappa_{a'}(T)$  and $\kappa_{b'}(T)$ 
were used. The best fits are 
shown in Fig.~\ref{Kappa} as solid lines and the corresponding fit parameters are 
listed in Table~\ref{t_1}. 

\subsection{Evaluation of the heat transport in the spin system}
\begin{figure}[t]
 \begin{center}
  \leavevmode
  \epsfxsize=0.9\columnwidth \epsfbox {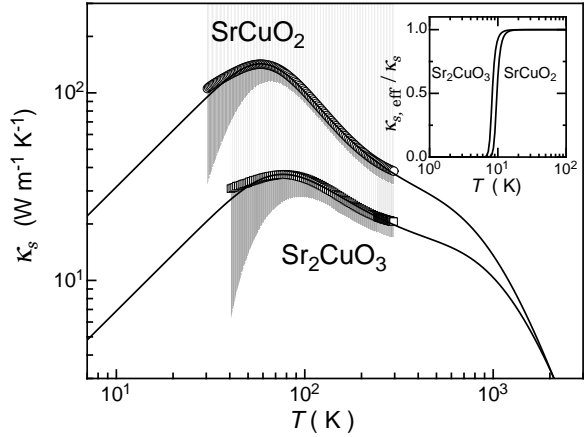}
  \caption{
  Spinon thermal conductivity for SrCuO$_{2}$ and Sr$_{2}$CuO$_{3}$. The 
  solid lines are calculated using Eqs.~(\ref{eKappaS}) and (\ref{eSpinMFP}). 
  The shaded areas 
  demonstrate possible errors caused by the uncertainty of the estimate 
  of the phonon thermal conductivity.
  The inset shows the ratios between the effective and the intrinsic spinon thermal 
  conductivities, calculated using Eq.~(\ref{eKappaEff}). 
  }
  \label{KappaS}
 \end{center}
\end{figure}

The spin contribution $\kappa_{s}$ to the heat transport along the chain 
direction was extracted by subtracting the calculated phonon contribution 
from the experimental data 
of $\kappa_{c'}(T)$. The resulting $\kappa_{s}(T)$ dependences are shown 
in Fig.~\ref{KappaS} as open symbols. It is these data that are 
used for the subsequent analysis of spinon heat transport. 
However, the way of evaluating $\kappa_{s}$, described above,  strongly 
relies on the assumption of isotropic phonon-phonon scattering. 
Alternatively, one may consider 
that the anisotropy of the phonon thermal 
conductivity does not completely vanish for $T > T_{\rm max}$, but
is similar in magnitude as for the case of chain materials Hg$_{2}$I$_{2}$ 
and Te discussed above (see the 
insert of Fig.~\ref{KAnisotropy}). 
The limiting case is then that the anisotropy of the phonon thermal conductivity is constant and 
equal to that in the region of $T_{\rm max}$.  
This situation is illustrated in Fig.~\ref{Kappa} where the dashed line 
represents the phonon thermal conductivity along the $c'$-axis, 
averaged  along the $a'$ and $b'$ -axes and 
scaled to the value of $\kappa_{c'}(T_{\rm max})$.
The result of subtracting this calculated phonon contribution from 
the experimental $\kappa_{c'}(T)$ values fixes the lower boundary 
of possible values of $\kappa_{s}(T)$ and coincides with the lower 
edge of the shaded areas in Fig.~\ref{KappaS}. 
These areas indicate the possible uncertainty of $\kappa_{s}(T)$,
caused by the ambiguity of our evaluation of the phonon parts of the 
thermal conductivities. The uncertainty is large at low temperatures, but 
at higher temperatures it is small enough to allow for a quantitative 
analysis of the thermal conductivity via spin excitations.

\section{Analysis of the spin contribution}
\subsection{Heat transport via diffusion}
An important question which can be answered immediately is whether the observed energy transport 
via spin excitations is diffusive or ballistic.  
For uniform 1D spin systems the energy transport is expected to be diffusive for 
spin $S>$1/2 
and ballistic for $S$=1/2.\cite{Huber69p602} However, the diffusive energy 
transport in $S$=1/2 chains may be restored by, for example, 
interchain interactions or intrachain next-nearest-neighbor 
interactions.\cite{Krueger71} 
In case of diffusive energy transport, the thermal conductivity is given by 
$\kappa_{s} = D_E C_{s}$, 
where $C_{s}$ is the spin specific heat per unit volume and $D_E$ is 
the energy diffusion constant. 
In the high-temperature limit, $D_{E}$ for a quantum spin system is given by 
$D_E = K J [S (S+1) ]^{1/2} a^2 / \hbar$, 
where $a$ is the distance between neighboring spins and $K$ is a 
constant which,  
in different theoretical models, varies between 0.25 and 
3.\cite{Huber69p534} 
In the low-temperature region, $D_E$ is  expected to be nearly 
independent of temperature.\cite{Huber68} 
Using, for the specific  heat, the low-temperature ($T \ll J/k_{B}$)  result of 
a Bethe ansatz solution\cite{Takahashi73} for a uniform chain of $N$ spins 
$S=1/2$
\begin{equation}\label{eLTCs}
 C_s =  { {2Nk_{B}^{2}} \over {3J}} T ,
\end{equation} 
the thermal conductivity 
in the diffusive regime is  
\begin{equation}\label{eKappaDiffusive}
 \kappa_{s} \sim  n_s  {{a^2 k_B^2 }\over{\hbar}} T.
\end{equation}
Here, $n_s$ is the number of spins per unit volume. 
Eq.~(\ref{eKappaDiffusive}) predicts rather small values for 
$\kappa_{s}$ (of the order of 1~W~(m~K)$^{-1}$ at 300~K) decreasing 
linearly with decreasing temperatures.  
The observed $\kappa_{s}(T)$ along the $c'$-axis is much higher than predicted by 
Eq.~(\ref{eKappaDiffusive}) and, at least at $T > 80$~K, it 
increases with decreasing temperature. This is a clear sign of 
non-diffusive energy transport via spin excitations along the chains, 
in agreement with theoretical predictions. It is worth noting that 
also the spin diffusion constant $D_{S}$ has experimentally been found to 
be strongly enhanced in Sr$_{2}$CuO$_{3}$.\cite{Takigawa96} 
This observation was considered 
as indicating the ballistic nature of spin transport. 

In contrast to the $c'$-direction, the high-temperature 
excess thermal conductivity along the $a'$ and $b'$ 
directions, discussed in section \ref{KappaPhab}, is in fair agreement 
with Eq.~(\ref{eKappaDiffusive}) both in absolute values and 
with respect to its temperature dependence. 
We therefore argue that the possible energy transport perpendicular to the Cu-O chains 
via the spin system
is small and diffusive. On the contrary, the energy 
transport via spins along the chains is substantial and relies on the 
ballistic propagation 
of spin excitations, interacting with lattice imperfections and other 
quasiparticles.
 
\subsection{Spinon heat transport along the chain direction}
\begin{figure}[t]
 \begin{center}
  \leavevmode
  \epsfxsize=0.9\columnwidth \epsfbox {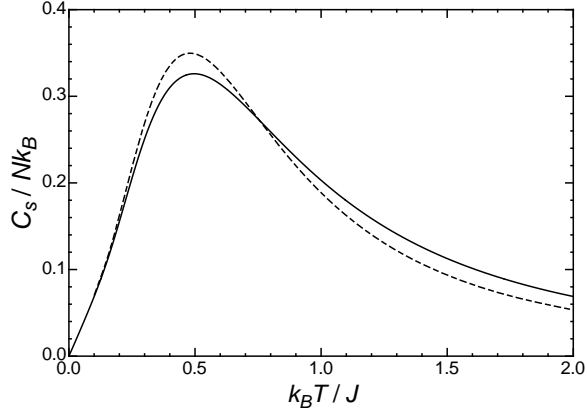}
   \caption{
   The  specific heat of the S=1/2 AFM Heisenberg chain, 
   calculated using Eq.~(\ref{eSpecificHeatSpinonsGeneral}) (solid line)
   and the Bethe ansatz  solution (dashed line, data from 
   Ref.~\protect\onlinecite{Kluemper2000}). 
  }
\label{CSpinons} 
\end{center}
\end{figure}

As mentioned in the introduction, 
the elementary excitations of a uniform Heisenberg 1D AFM $S=1/2$ system  
carry a spin $S=1/2$.\cite{Faddeev81} Therefore
the low-temperature thermodynamic behavior of such an ensemble can be 
accounted for 
by using the usual Fermi-Dirac distribution $f = (\exp (\varepsilon /k_B T)+1)^{-1}$
with zero chemical potential,\cite{McRae98} where 
$\varepsilon$ is the spinon energy.
The dispersion relation for spinons is given by\cite{Faddeev81} 
\begin{equation}\label{eSpinonDisp}
  \varepsilon (k)={{J\pi } \over 2}\left| {\sin ka} \right|.
\end{equation} 
For spinons, the possible values of the wave vector $k$ are restricted 
to one half of the Brillouin zone. 
The low-temperature specific heat, calculated using the 
general equation  
\begin{equation}\label{eSpecificHeatSpinonsGeneral}
C_s=   { 2Na \over {\pi}}    \int \limits_{0}^{\pi /2a} 
            {  {\partial f} \over  {\partial T} }
	    \varepsilon   {dk},
\end{equation} 
agrees with the low-temperature  result of Eq.~(\ref{eLTCs}). 
As illustrated in Fig.~\ref{CSpinons}, the high-temperature specific heat   
calculated with Eq.~(\ref{eSpecificHeatSpinonsGeneral}) 
qualitatively reproduces the main high-temperature features of the results of 
the Bethe ansatz solution.\cite{Bloete74,Kluemper2000} The 
disagreement between the two models is not important for our analysis 
since our results have been obtained 
in the  low temperature region $k_{\rm B} T /J < 0.15$.   

Applying the Boltzmann-type approximation, the 
thermal conductivity of a spinon system is 
\begin{equation}\label{eKappaSpinonsGeneral}
  \kappa_s =   {2 n_s a \over {\pi}}    \int \limits_{0}^{\pi /2a} 
            {  {\partial f} \over  {\partial T} }
	    \varepsilon v_s l_s {dk},
\end{equation} 
where $v_s=\hbar^{-1} {\partial \varepsilon} / {\partial k}$ is the 
group velocity and $l_s$ is the mean free path of a spinon.
Cast in another but equivalent form, 
\begin{equation}\label{eKappaS}
    \kappa_s =  \frac {2 n_s k_{\rm B}^{2} a  } { \pi \hbar } T      
    \int\limits_{0}^{J \pi/ 2 k_{\rm B} T}
    \frac{x^{2}e^{x}}{(e^{x}+1)^{2}}l_{s}(\varepsilon,T) dx,
\end{equation}
where $x=\varepsilon/ k_{\rm B} T$.
If several independent scattering mechanisms have to be considered, the spinon  mean free 
path can be represented as 
$l_{s}^{-1}(\varepsilon,T)=\sum l_{s,i}^{-1}(\varepsilon,T)$, 
where each $l_{s,i}(\varepsilon,T)$ term corresponds to an independent scattering 
channel. This is analogous to the case of phonons, considered in  Eq.~(\ref{eTau}).  

\begin{figure}[t]
 \begin{center}
  \leavevmode
  \epsfxsize=0.9\columnwidth \epsfbox {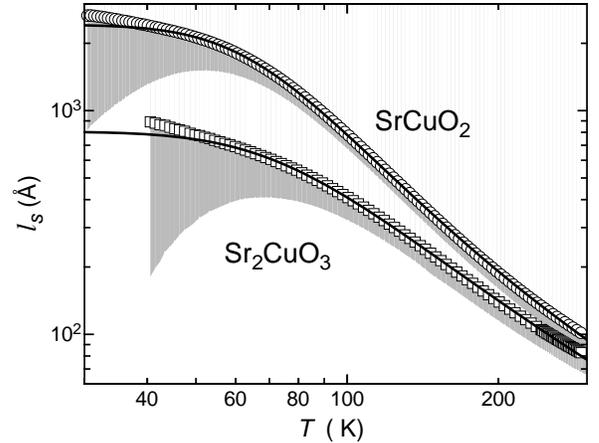}
  \caption{
  The temperature dependences of the spinon mean free paths for 
  Sr$_{2}$CuO$_{3}$ and SrCuO$_{2}$, calculated using Eq.~(\ref{eKappaS}). 
  The solid lines are 
  fits using Eq.~(\ref{eSpinMFP}). The shaded areas have the same 
  meaning as in Fig.~\ref{KappaS}.
  }
  \label{MagMFP}
 \end{center}
\end{figure}

The most straightforward way to analyze our $\kappa_{s}(T)$ data  would be 
analogous to that employed for the phonon conductivity, i.e., by choosing a set of 
$l_{s,i}(\varepsilon,T)$ terms in the equation for the mean free path of spinons and 
fitting the data of $\kappa_{s}(T)$ to Eq.~(\ref{eKappaS}). However, 
the energy and temperature dependences for most of the plausible scattering 
mechanisms have not yet  been worked out. 
That is why we treated the $\kappa_{s}(T)$ data  
by inserting the spinon mean free path $l_s(T)$, averaged 
over all $\varepsilon$, at a given temperature.
In that case, $l_s(T)$ may be taken out of the integral in 
Eqs.~(\ref{eKappaSpinonsGeneral}) and (\ref{eKappaS}) 
and thus may be calculated from the $\kappa_{s}(T)$ data, using the known 
values of $J$ for the materials investigated here. 
For these calculations, $J/k_{\rm B} =$  2100~K (Ref.~\onlinecite{Motoyama96}) 
and  2620~K 
(Ref.~\onlinecite{Sologubenko2000_213})  were used 
for  SrCuO$_{2}$ and Sr$_{2}$CuO$_{3}$, respectively.
The resulting values of 
$l_s(T)$ for SrCuO$_{2}$ and  Sr$_2$CuO$_3$ are shown in 
Fig.~\ref{MagMFP}.

Here again, the shaded areas reveal the possible range of $l_s$  
values caused by the 
uncertainties   in evaluating the phonon thermal conductivities. 
In Ref.~\onlinecite{Sologubenko2000_213}, it was noted that $l_s(T)$ 
for Sr$_{2}$CuO$_{3}$ may be approximated by  
\begin{equation}\label{eSpinMFP}
l_s^{-1} = l_{sp}^{-1} + l_{sd}^{-1} = A_{sp} T \exp(-T^*/T) +  L_{sd}^{-1},
\end{equation}
where the two terms were attributed to spinon scattering by phonons 
and defects, respectively. Here, the parameter  $A_{sp}$   
characterizes the strength of spin-lattice interaction, $T^*$ is 
related to the minimum energy needed to produce a single spinon-phonon 
Umklapp-process, and $L_{sd}$ is the mean distance between 
magnetic defects effectively scattering spin excitations in a spin 
chain.
It may be seen that with Eq.~(\ref{eSpinMFP}) and the appropriate 
parameters, 
$l_s(T)$ may be well approximated not only for Sr$_{2}$CuO$_{3}$ but also 
for SrCuO$_{2}$.
The  values of 
$A_{sp}$=7.1$\times10^{5}$~m$^{-1}$K$^{-1}$, $T^{*}$=177~K,  
$L_{sd}$=8.1$\times10^{-8}$~m for Sr$_{2}$CuO$_{3}$, 
and 
$A_{sp}$=6.7$\times10^{5}$~m$^{-1}$K$^{-1}$, $T^{*}$=204~K, 
$L_{sd}$=2.42$\times10^{-7}$~m for SrCuO$_{2}$,
were obtained by fitting the data of $l_s(T)$ with Eq.~(\ref{eSpinMFP}).
The fits are shown as solid lines in Fig.~\ref{MagMFP}.
The parameter values for Sr$_{2}$CuO$_{3}$ are slightly different from 
those presented in Ref.~\onlinecite{Sologubenko2000_213}, because of a
somewhat different way of evaluating the phonon thermal conductivity.
The agreement between the experimental data and the fit is very good 
above approximately 50~K, but at lower temperatures the uncertainty in 
extracting $l_s(T)$ becomes far too large to draw any reliable 
conclusion concerning
its temperature dependence  and the involved scattering mechanisms.

The fit values of the parameters $A_{sp}$ and $T^{*}$ which are
related to the spin-lattice interaction 
are similar in magnitude for the two compounds, in spite of 
the different chain arrangements. 
The values of $T^*$ scale with the Debye temperatures, 
supporting the idea that the term $l_{sp}^{-1}$ in 
Eq.~(\ref{eSpinMFP}) is indeed related to the spin-lattice interaction. 
It appears that the main difference between the two materials 
is in the defect scattering, which is three times 
more efficient for Sr$_{2}$CuO$_{3}$ than for SrCuO$_{2}$. The 
low-temperature values of $l_s$ for  Sr$_{2}$CuO$_{3}$ have been 
found\cite{Sologubenko2000_213} to be in fair agreement with 
the mean distance between 
neighboring bond-defects causing local changes in the 
magnetic coupling between Cu spins, as extracted from NMR data.\cite{Boucher2000}
Unfortunately, similar NMR results revealing the distance 
between bond-defects are absent for SrCuO$_{2}$.  Our data suggest 
an about three times smaller concentration of these defects in 
SrCuO$_2$ than in  Sr$_{2}$CuO$_{3}$.

\subsection{Influence of spinon-phonon scattering}

For the  two investigated materials, the evaluation of the spinon part  of the 
thermal conductivity has been achieved only in a limited temperature region, 
characterized by $0.015 \leq T/J \leq 0.15$. Two mechanisms of spinon 
scattering seem to be effective, 
namely the temperature-dependent spinon-phonon interaction 
and  the temperature-independent scattering of spinons by defects. 
The significance of other scattering mechanisms, if 
any, is negligible. For example, normal processes of spinon-spinon 
scattering cannot change the total momentum of the spinon system and, 
therefore, should not contribute to the thermal resistance. The 
spinon-spinon Umklapp-processes are expected to play a serious role
only at temperatures  $T \sim J/k_{\rm B}$. 

Supposing that the validity of 
Eq.~(\ref{eSpinMFP}) extends  
beyond the region  $0.015 \leq 
T/J \leq 0.15$, we may estimate the expected spinon thermal conductivity 
at lower and higher temperatures 
using Eq.~(\ref{eKappaS}). The result is shown as solid lines in Fig.~\ref{KappaS}.      
Since $l_s(T)$ is constant and equal to $L_{sd}$ at very low temperatures,  
$\kappa_{s}(T)$  is expected to vary linearly with $T$ as 
$  \kappa_{s} =  n_s a k_{\rm B}^{2}\pi L_{sd} T /3 \hbar$.
The phonon thermal conductivity 
varies as $T^{3}$  at low temperatures and hence 
the spinon contribution should progressively dominate with decreasing temperature.
This expectation is obviously not met by our low-temperature results. 
First of all, no linear contribution to the total measured thermal 
conductivity has been identified and, second, no distinct feature in 
$\kappa(T)$  is observed at the N\'{e}el 
temperature $T_{N}$, 
indicating a negligible spinon contribution $\kappa_s$. This suggests 
that at some temperature below 30~K the spinon contribution  deviates from the  
predicted behavior shown in Fig.~\ref{KappaS} and by  
6~K, spinons are excluded from heat transport. 
At this point, a discussion of the spin-lattice interaction is 
in order. 

Although the spin-lattice interaction reduces both the phonon- and 
the spin-related thermal conductivities because of scattering 
processes involving both types of quasiparticles, 
some degree of it is needed for the spin-related heat conduction to be 
observable in a thermal conductivity experiment.\cite{Sanders77} 
The energy  provided by a heater generates 
only phonons, and the spin-phonon interaction is needed for an energy 
transfer from the lattice to  
the spin system.  The effective spinon thermal conductivity $\kappa_{s,\rm eff}$ 
which is accessible in a coupled spin-lattice system, is\cite{Sanders77}
\begin{equation}\label{eKappaEff}
  \kappa_{s,\rm eff}= (\kappa_{s} + \kappa_{ph}) 
 \left( {1+{{\kappa_s} \over {\kappa_{ph}}}{{\tanh (A L_{\rm sample}/2)} \over 
 {A  L_{\rm sample}/2}}} \right)^{-1}        - \kappa_{\rm ph},
\end{equation}
where $L_{\rm sample}$ is the sample length,  
\begin{equation}\label{eA}
    A  = \left(  \tau_{sp}^{-1}  { { \kappa_{s}^{-1} + 
    \kappa_{ph}^{-1} } \over {  C_{ph}^{-1} + C_s^{-1} } } \right)^{1/2},
\end{equation}
and $\tau_{sp}$ is the spin-lattice relaxation time. For very short 
spin-lattice relaxation times ($\tau_{sp}^{-1} \to \infty$), 
$\kappa_{s, \rm eff}=  \kappa_{s}$. 
In the opposite case of a very weak spin-lattice 
interaction ($\tau_{sp}^{-1} \to 0$), however, $\kappa_{s,\rm eff}= 0$ and 
the thermal transport by spinons cannot be observed in an experiment, 
regardless of how large the intrinsic $\kappa_{s}$ of the sample is. 

If the first term in Eq.~(\ref{eSpinMFP}) is indeed dictated by the 
spin-lattice interaction, it is possible to establish 
the temperature region in which the observed $\kappa_{s,{\rm eff}}$ 
equals the intrinsic $\kappa_{s}$. 
For this purpose, the ratio 
$\kappa_{s,{\rm eff}}/\kappa_{s} $ 
was calculated by assuming $\tau_{sp}= l_{sp}/v_{s}$ (see Eq.~(\ref{eSpinMFP}) )
and  $v_s = J a \pi /2\hbar$, valid for low-energy spinons 
which dominate at $T \ll J/k_{B}$. 
The result is shown in the inset of Fig.~\ref{KappaS}.
Although the extrapolation of 
Eq.~(\ref{eSpinMFP}) to lower temperatures 
may  not exactly be correct, it is nevertheless clear (see the inset 
of Fig.~\ref{KappaS}) that at very low 
temperatures the spin-lattice interaction turns out to be too weak 
for the spinon thermal conductivity to be observed. 

From our discussion it is obvious that the interaction between the spin 
and the phonon  system should 
show up both in spinon and phonon heat transport. However, 
other mechanisms of phonon scattering, mainly the
phonon-phonon interaction, may be stronger  
and mask the scattering of phonons by spin excitations. 
This may be the reason why a term 
corresponding to the phonon-spinon scattering  does not have to be 
included in Eq.~(\ref{eTau}) for a reasonable description of 
$\kappa_{\rm ph}(T)$.
On the other hand, a phonon-spinon relaxation rate 
$\bar \tau_{ps}(T)$, averaged over the entire 
phonon spectrum at a given temperature, may  
roughly be approximated by 
\begin{equation}\label{eTaups_Lsp}
   {\bar \tau_{ps}^{-1}} \sim {{N_s} \over {N_{\rm ph}}} \tau_{sp}^{-1},   
\end{equation}
where at a fixed temperature $T$, $N_{\rm ph}(T)$ and $N_s(T)$ are 
the average numbers of phonons and spinons per unit volume, respectively.
At high temperatures, where spin-phonon 
interaction is  important, Eq.~(\ref{eTaups_Lsp})  for the
phonon-spinon scattering rate gives
${\bar \tau_{ps}^{-1}} \propto T \exp(-T^*/T)$, i.e., the same 
temperature dependence as is characteristic for the
phonon-phonon interaction. That is why, even if the influence of phonon 
scattering by spin excitations is not negligible, it is difficult to 
separate it from the phonon-phonon interaction. 

In the present discussion, we do not consider the possibility 
of a mutual drag between spin and lattice excitations, 
which may be considerable under certain conditions, 
as described in Ref.~\onlinecite{Gurevich67}. 

\section{Summary}

In this paper the thermal conductivities of the spin-chain compounds  SrCuO$_{2}$ 
and  Sr$_{2}$CuO$_{3}$ were studied. Although the crystal structures of the two 
compounds are different in the sense that the former  contains linear  
and the latter zig-zag Cu-O chains, the 
thermal conductivity of both materials is remarkably similar. In 
particular, the heat transport in the directions perpendicular to the   
chains is dominated by phonons, but along the chain direction 
and at high temperatures there is a substantial excess contribution related to 
the transport of energy  by spinons. 

The phonon thermal conductivity is analyzed employing a Debye-type 
approximation. The main sources of phonon scattering are 
phonons at high temperatures and lattice defects, presumably 
dislocations, at low temperatures. The spin-phonon interaction  is 
not seen in the phonon heat transport, most likely because it is masked by 
other scattering processes.

An eventual energy transport via the spin system in the directions perpendicular 
to the spin chains is found to be small and, if any at all, of diffusive character. 
On the contrary, the spin-mediated energy transport along the chain 
direction is large and propagating ballistically.    
The  contribution of spin excitations (spinons) 
is reliably singled out  along the chain 
direction at high temperatures where it exceeds the phonon 
contribution quite substantially. Our analysis  based on a fermion model suggests 
that the expected infinite spinon thermal conductivity is limited by 
the influence of defects and phonons.  

\acknowledgments
This work was financially supported in part by
the Schweizerische Nationalfonds zur F\"{o}rderung der Wissenschaftlichen
Forschung.

\pagebreak
\onecolumn
\begin{table}[hbp]
\caption{Parameters of the fitting of $\kappa(T)$ data to Eq. 
 (\ref{eLambda}).}
\label{t_1}
\begin{tabular}{lcccccc}
\multicolumn{1}{c}{Parameter} &
\multicolumn{1}{c}{SrCuO$_{2}$} &
\multicolumn{1}{c}{SrCuO$_{2}$} &
\multicolumn{1}{c}{Sr$_{2}$CuO$_{3}$} &
\multicolumn{1}{c}{Sr$_{2}$CuO$_{3}$} &
\multicolumn{1}{c}{SrCuO$_{2}$} &
\multicolumn{1}{c}{Sr$_{2}$CuO$_{3}$} \\
\multicolumn{1}{c}{} &
\multicolumn{1}{c}{$a'$-axis} &
\multicolumn{1}{c}{$b'$-axis} &
\multicolumn{1}{c}{$a'$-axis} &
\multicolumn{1}{c}{$b'$-axis} &
\multicolumn{1}{c}{$c'$-axis} &
\multicolumn{1}{c}{$c'$-axis} \\    
\tableline 
$L$ ($10^{-3}$~m)      & 0.32$\pm$0.1   & 0.25$\pm$0.02 &   0.97$\pm$0.5  
                       &  8$\pm$6   &  0.63$\pm$0.04   &    2.5$\pm$2  \\ 
$A$ ($10^{-42}$ s$^{3}$)      & $<10^{-45}$     & $<10^{-45}$  & $<10^{-45}$     
                              &$<10^{-45}$  &$<10^{-45}$  & $<10^{-45}$  \\ 
$B$ ($10^{-18}$ s K$^{-1}$)  & 11.1$\pm$0.5  & 13.5$\pm$0.5  &  7.6$\pm$0.5
                             &  9.0$\pm$0.5 & 12.3  & 8.30  \\ 
$b$                    & 2.7$\pm$0.2   & 2.8$\pm$0.1  &  3.5$\pm$0.3 
                       & 3.3$\pm$0.2    & 2.75            &  3.4   \\
$C$  ($10^{-6}$)        & 7.2$\pm$1   &   5.8$\pm$0.3  &  9.4$\pm$1  
                        & 10.6$\pm$0.7   & 3.9$\pm$0.2  &  5.5$\pm$0.2   \\
$D$ ($10^{9}$ s$^{-1}$)  & 4.6$\pm$2   & 2.7$\pm$1  & 5.6$\pm$3  
                         & 6.5$\pm$2.5   & 3.5$\pm$1 & 7.7$\pm$0.2  \\
$\omega_{0}$ ($10^{12}$ s$^{-1}$) & 10.0$\pm$1 & 8.3$\pm$1.5 & 10.2$\pm$1
                                  & 9.6$\pm$1  &   9.2       & 9.9  \\
$c$              &  0.86$\pm$0.03       & 0.90$\pm$0.01       &  0.76$\pm$0.03 
                       & 0.87$\pm$0.03 &  0.88  &  0.815 \\
Temperature region     &  $T<$100~K   &	$T<$100~K &   $T<$100~K   
                       &  $T<$100~K   & $T<$12~K & $T<$13~K \\
\end{tabular}
\end{table}


\begin{references}
\bibitem{_Comment1}
The fact that the absolute temperatures of the anomalies are not exactly the
  same for $\beta(T)$ and $\chi''_{c}(\omega,T)$ should not be overrated
  because the temperatures of the $\chi''_{c}(\omega,T)$ peaks are
  frequency-dependent and shift to lower temperatures for lower $\omega$
  (Ref.~\onlinecite{Zhai99}).
\end{references}
\end{document}